\documentclass[aps,prd,twocolumn,showpacs,floatfix,amsmath,amssymb,nofootinbib,longbibliography]{revtex4-1}

\usepackage{graphicx}
\usepackage[hidelinks]{hyperref}
\usepackage{bm}
\usepackage{bbold}
\usepackage{xcolor}

\hypersetup{colorlinks,
   linkcolor={red!50!black},
   citecolor={blue!50!black},
   urlcolor={blue!75!black}}

\allowdisplaybreaks

\graphicspath{{.}{./figs/}}

\newcommand{\be}{\begin{equation}}
\newcommand{\ee}{\end{equation}}
\newcommand{\ba}{\begin{eqnarray}}
\newcommand{\ea}{\end{eqnarray}}

\def\nuc#1#2{\relax\ifmmode{}^{#1}{\protect\text{#2}}\else${}^{#1}$#2\fi}
\newcommand{\nm}{\ensuremath{N_\mathrm{max}}}
\newcommand{\NN}{\ensuremath{N\!N}}
\newcommand{\NNN}{\ensuremath{N\!N\!N}}
\newcommand{\Amo}{A\text{-}1}
\newcommand{\ket}[1]{\lvert #1 \rangle}

\allowdisplaybreaks[4]

\begin{document}

\title{\textit{Ab initio} nuclear response functions for dark matter searches} 

\author{D.~Gazda} 
\affiliation{Department of Physics,
  Chalmers University of Technology, SE-412 96 G\"oteborg, Sweden}

\author{R.~Catena} 
\affiliation{Department of Physics,
  Chalmers University of Technology, SE-412 96 G\"oteborg, Sweden}

\author{C.~Forss\'en} 
\affiliation{Department of Physics,
  Chalmers University of Technology, SE-412 96 G\"oteborg, Sweden}

\begin{abstract}
  \noindent
  We study the process of dark matter particles scattering off \nuc{3,4}{He}
  with nuclear wave functions computed using an \textit{ab initio} many-body
  framework. We employ realistic nuclear interactions derived from chiral
  effective field theory at next-to-next-to-leading order (NNLO) and develop an
  \textit{ab initio} scheme to compute a general set of different nuclear
  response functions. In particular, we then perform an accompanying uncertainty
  quantification on these quantities and study error propagation to physical
  observables. We find a rich structure of allowed nuclear responses with
  significant uncertainties for certain spin-dependent interactions. The
  approach and results that are presented here establish a new framework for
  nuclear structure calculations and uncertainty quantification in the context
  of direct and (certain) indirect searches for dark matter.
\end{abstract}

\maketitle 

\section{Introduction}
\noindent Convincing evidence for the presence of dark matter in the Universe
has been gathered over the past decades~\cite{Bertone:2016nfn}. In the standard
paradigm of modern cosmology, dark matter is a weakly interacting massive
particle (WIMP)~\cite{Jungman:1995df,Bergstrom00,Bertone:2004pz}. On
cosmological scales, it behaves like a dissipationless and nonrelativistic
fluid from the beginning of cosmological structure formation until present
time~\cite{Frenk:2012ph}. The detection of dark matter particles in a laboratory
or in space is currently a priority of astroparticle physics. The experimental
technique known as direct detection is expected to play a key role in this
context~\cite{Baudis:2012ig}. It searches for nuclear recoil events induced by
the nonrelativistic scattering of Milky Way dark matter particles in
low-background detectors. Reliable nuclear physics input is therefore needed for
the interpretation of data from such experiments. However, there is currently a
gap between the treatment of the nuclear physics input in the field of dark
matter studies and the level of sophistication that has been reached in modern
theoretical nuclear physics. It is the main purpose of this work to fill this
gap and introduce a more systematic approach, with a solid theoretical
underpinning, that also allows to explore and quantify various sources of
uncertainties. The frameworks that will be used to achieve this goal are
effective field theories (EFTs) and nuclear \textit{ab initio} many-body
methods.

Effective theory methods have already proven to be a very powerful tool in the
analysis of dark matter detection
experiments~\cite{Chang:2009yt,Fan:2010gt,Fitzpatrick:2012ix}. The main
advantage of the effective theory approach to dark matter is that it allows for
a model-independent analysis of available data. In contrast, signs of important
physical properties can be obscured when using a simplistic model for dark
matter interactions. At the same time spurious correlations among physical
observables can be enforced through an inappropriately small number of model
parameters. Two main approaches have been used when constructing an effective
theory for WIMP--nucleus scattering. In both cases the end result is an EFT with
nonrelativistic nucleon and WIMP fields as the relevant degrees of freedom.
Firstly, one can consider a specific set of effective interaction terms at the
quark level defined at the hadronic scale and use chiral symmetry constraints to
estimate the hierarchy among one- and two-nucleon currents
~\cite{Cirigliano:2012pq,Menendez:2012tm,Klos:2013rwa,Klos:2014ck,Hoferichter:2015ipa,PhysRevD.94.063505}.
This approach is very appealing since similar constraints are used in the
construction of nuclear
forces~\cite{vanKolck:1994yi,Epelbaum:2008ga,Machleidt:2011zz}. However, the
matching of standard model fields to hadronic-level operators is an intricate
problem; see e.g.\ Ref.~\cite{Beane:2013kca} with lattice QCD results and a
brief discussion of potential power-counting issues that are relevant for
meson-exchange current contributions to the WIMP--nucleus interaction. The
chiral EFT approach also allows, in principle, a mapping to the parameter space
of new-physics models~\cite{PhysRevD.94.063505}. Such a mapping is not
straightforward and requires us to take into account the evolution of the WIMP
effective operators from the mediator mass scale to the low-energy hadronic
scale that is probed by direct detection experiments \cite{DEramo:2014nmf}.

Alternatively, one can integrate out the QCD dynamics and construct directly an
EFT in which nonrelativistic nucleon and WIMP fields are the degrees of
freedom~\cite{Fitzpatrick:2012ix}. This so-called NREFT generates the most
general set of WIMP--nucleon interactions based only on the requirement of
Galilean invariance and momentum conservation. In this approach the connection
to symmetries of QCD is lost, but can be recovered by matching to the chiral EFT
framework discussed above.  In this context we mention explicitly the work by
\textcite{Hoferichter:2015ipa} to present a common chiral power-counting scheme
and to match it to the NREFT operator basis. Most importantly, however, the
NREFT framework is less restrictive with respect to the assumptions on the
underlying quark-level dark matter interactions and the type of dark matter
particle. There are possible dark matter--quark interactions, such as e.g.\ a
dimension-6 tensor--tensor type~\cite{DelNobile:2013sia}, that have not yet been
analyzed in the chiral EFT framework and therefore may alter some of its
conclusions. At the moment there is no experimental evidence that would favor
any particular form of the underlying dark-matter particle interactions and we
will therefore work within the general NREFT framework. We stress, however, that
the \textit{ab initio} nuclear-physics method that we present here can be
applied to both EFT approaches as its starting point is the interaction at the
nucleon level.

The construction of the dark matter--nucleus interaction is the next step in the
effective-theory approach to dark matter detection. Following
\textcite{Fitzpatrick:2012ix}, eight nuclear response functions can be generated
in the dark matter elastic scattering by nuclei. The interpretation of any dark
matter experiment probing the dark matter--nucleus interaction is unavoidably
affected by the uncertainties within which these nuclear response functions are
known. In the simplest treatment, only spin-independent interactions are
considered and phenomenological nuclear response functions--- so-called `Helm
form factors'---are used. More recently, many additional responses have been
considered and more sophisticated nuclear-structure calculations have been
performed using the shell model
(SM)~\cite{Engel:1992bf,PhysRevC.79.044302,Menendez:2012tm,Anand:2013yka,Klos:2013rwa,Klos:2014ck,Catena:2015uha}.
The SM is arguably a very successful phenomenological model for nuclear
structure, see e.g.\ Refs.~\cite{Brown:2001zz,RevModPhys.77.427} for general
overviews. Its configuration space comprises a relatively small number of
``active'' particles outside a core of nucleons that are frozen in the
lowest-energy orbitals and not included in the calculation. This significant
truncation of the model space is often critical for allowing any kind of
solution to the many-body problem. The residual valence-space interaction
should, in principle, incorporate effects from degrees of freedom that are not
explicitly included in the model space. In practice, its construction typically
corresponds to the introduction of free fitting parameters that are tuned so
that the model reproduces (with an acceptable accuracy) energy spectra and/or
other observables in the region of interest. The ability to quantify theoretical
uncertainties associated with predicted nuclear matrix elements, and
consequently in the constructed form factors, becomes severely restricted in
such an approach. Only very recently it was shown how to obtain residual
effective valence-space interactions starting from the underlying microscopic
internucleon interaction in a systematic, nonperturbative framework using
\textit{ab initio} methods
\cite{PhysRevLett.113.142502,PhysRevLett.113.142501,Dikmen:2015tla}. However, it
remains to be studied how theoretical model uncertainties can be quantified.

In recent years, \textit{ab initio}
methods~\cite{2013PrPNP..69..131B,Hagen:2014jt,Carlson:2015fu,Hergert:2016fd,Lee:2016fhn,Barbieri:2016uib}
have matured to a level where precise nuclear many-body calculations can be
performed starting from nucleons as the relevant degrees of freedom and using
realistic internucleon interactions. Furthermore, the use of EFTs for the
description of these nuclear interactions provides a systematic approach that
offers an estimate of the inherent model error. Significant progress in the
quantification of truncation errors in EFT was reported
recently~\cite{Epelbaum:2015,PhysRevC.92.024005,Carlsson:2016cw} and also
employed to provide theoretical uncertainties in nuclear structure
calculations~\cite{Carlsson:2016cw,Binder2016} by combining \textit{ab initio}
many-body methods and chiral EFT interactions. It is a specific aim of this work
to demonstrate how nuclear uncertainties can be quantified, at least for
selected proof-of-principle cases, using \textit{ab initio} methods and
realistic internucleon interactions.

Experiments whose analysis are affected by these uncertainties are dark matter
direct detection experiments, with various detector materials, and neutrino
telescopes searching for neutrinos from dark matter annihilations in the Sun and
the Earth. In this work we concentrate on the former ones, and in particular on
detector designs with sensitivities to the direction of nuclear recoils. Such
designs are currently in a research and development stage. They are of
particular interest for the present analysis in that helium, and especially
\nuc{3}{He}, is one of the target materials explored in this
context~\cite{Ahlen:2010,Mayet:2016zxu}. For such a light target nucleus,
\textit{ab initio} nuclear structure calculations are straightforward, which
allows a more robust uncertainty quantification. Furthermore, the use of
\nuc{3}{He} for dark matter detection is interesting for other
reasons~\cite{Moulin:2003br,Moulin:2005fz,Franarin:2016ppr}: it is an ideal
target for the detection of light dark matter particles; neutron rejection can
easily be achieved through the process $n+\nuc{3}{He} \rightarrow
p+\nuc{3}{H}+764$~keV; it has no intrinsic x-ray emission and a low natural
radioactive background; it can be polarized; and it allows to probe the spin of
the dark matter particle. As far as \nuc{4}{He} is concerned, its use for dark
matter detection has recently been considered in~\cite{Schutz:2016tid}. In this
investigation we will focus on hypothetical \nuc{3}{He} and \nuc{4}{He}
detectors with directional sensitivity.

The article is organized as follows. In Sec.~\ref{sec:methodology} we first
review the nonrelativistic effective theory of dark matter--nucleon interactions
(Sec.~\ref{sec:dmtheory}) and then introduce the \textit{ab initio} no-core shell
model technique for the calculation of nuclear matrix elements in a Jacobi,
relative-coordinate basis (Sec.~\ref{sec:abinitio}). The nuclear many-body problem is
solved with chiral nuclear interactions as input and these will be introduced in
Sec.~\ref{sec:chiral}. Results are presented in Sec.~\ref{sec:results}, focusing
on nuclear response functions in Sec.~\ref{sec:W} and on rates of dark
matter--nucleus scattering events at directional detectors in
Sec.~\ref{sec:dmpheno}. We conclude with an outlook in
Sec.~\ref{sec:conclusions}.

\section{Methodology}
\label{sec:methodology}

\subsection{Dark matter--nucleon and nucleus interaction}
\label{sec:dmtheory}
\noindent Consider the nonrelativistic scattering of a dark matter particle
$\chi$ by a single nucleon $N$:~$\chi(\mathbf{k}) + N(\mathbf{p}) \rightarrow
\chi(\mathbf{k}') + N(\mathbf{p}')$, where initial and final three-dimensional
momenta are denoted by $\mathbf{k}$ and $\mathbf{p}$, and $\mathbf{k}'$ and
$\mathbf{p}'$, respectively. Three-dimensional momentum conservation and
Galilean invariance, i.e., the invariance under constant shifts of particle
velocities, constrain the transition amplitude, $\mathcal{M}$, for this process.
Momentum conservation implies that only three of the four momenta $\mathbf{k}$,
$\mathbf{p}$, $\mathbf{k}'$ and $\mathbf{p}'$ are independent. The momentum
transfer $\mathbf{q}=\mathbf{k}-\mathbf{k}'$, $\mathbf{k}$ and $\mathbf{p}$ form
a possible set of independent momenta. Galilean invariance implies that
$\mathcal{M}$ cannot depend on $\mathbf{k}$ and $\mathbf{p}$ separately, but
only on a Galilean invariant combination of them, for instance $\mathbf{v}=
\mathbf{k}/m_\chi-\mathbf{p}/m_N$, where $m_\chi$ and $m_N$ are the dark matter
particle and nucleon mass, respectively. Here $\mathbf{v}$ is the dark
matter--nucleon relative velocity, and $\mathbf{q}$ is per se Galilean
invariant. We conclude that in general,
$\mathcal{M}=\mathcal{M}(\mathbf{q},\mathbf{v},\mathbf{S_\chi},\mathbf{S_N})$,
where $\mathbf{S_\chi}$ and $\mathbf{S_N}$ are the dark matter and nucleon spin,
respectively.

Next, we focus on the nonrelativistic quantum mechanical Hamiltonian
$\hat{\mathcal{H}}_{\chi N}$ underlying the scattering amplitude $\mathcal{M}$.
At the quantum mechanical level, any interaction operator describing the
nonrelativistic limit of dark matter--nucleon interactions can be expressed in
terms of four Hermitian
operators~\cite{Fitzpatrick:2012ix}:~$i\mathbf{\hat{q}}$, where
$\mathbf{\hat{q}}$ is the momentum transfer operator; the transverse relative
velocity operator; $\mathbf{\hat{v}}^{\perp}$, and the dark matter particle and
nucleon spin operators, $\mathbf{\hat{S}}_\chi$ and $\mathbf{\hat{S}}_N$,
respectively. By construction $\mathbf{\hat{v}}^{\perp} \cdot
\mathbf{\hat{q}}=0$. Without further restrictions, $\hat{\mathcal{H}}_{\chi N}$
can in principle include an infinite number of interaction operators:~all scalar
combinations of $i \mathbf{\hat{q}}$, $\mathbf{\hat{v}}^{\perp}$,
$\mathbf{\hat{S}}_\chi$ and $\mathbf{\hat{S}}_N$. However, when $|\mathbf{q}|$
is small compared to the mass of the particle that mediates the dark
matter--nucleon interaction, $\hat{\mathcal{H}}_{\chi N}$ can be expanded in
powers of ${\bf{\hat{q}}}$. Truncating the expansion at second order, only
14 independent Galilean invariant interaction operators arise if dark
matter has spin less than or equal to 1/2~\cite{Anand:2013yka}. For spin 1 dark
matter, two additional operators can be constructed~\cite{Dent:2015zpa},
although these are only relevant when specific operator interference patterns
are not negligible. We list the interaction operators $\hat{\mathcal{O}}_j$
considered in this study in Table~\ref{tab:operators}, using the notation
introduced in~\cite{Anand:2013yka} and an index $j$ to label them.

The operators in Table~\ref{tab:operators} define a nonrelativistic theory
called effective theory of dark matter--nucleon interactions (NREFT). Initially
formulated
in~\cite{Chang:2009yt,Fan:2010gt,Fitzpatrick:2012ix,Fitzpatrick:2012ib,Anand:2013yka},
it has later been developed
in~\cite{Menendez:2012tm,Cirigliano:2012pq,DelNobile:2013sia,Klos:2013rwa,Peter:2013aha,Hill:2013hoa,Catena:2014uqa,Catena:2014hla,Catena:2014epa,Gluscevic:2014vga,Panci:2014gga,Vietze:2014vsa,Barello:2014uda,Catena:2015uua,Schneck:2015eqa,Dent:2015zpa,Catena:2015vpa,Kavanagh:2015jma,D'Eramo:2016atc,Catena:2016hoj,Kahlhoefer:2016eds}.
In this context, the most general Hamiltonian density for nonrelativistic dark
matter--nucleus interactions is
\begin{equation}
\hat{\mathcal{H}}_{\chi A}= \sum_{i=1}^{A}  \sum_{\tau=0,1} \sum_{j} c_j^{\tau}\hat{\mathcal{O}}_{j}^{(i)} \, t^{\tau}_{(i)} \,,
\label{eq:H_chiT}
\end{equation}
where $A$ is the mass number of the target nucleus. The matrices
$t^0_{(i)}=\mathbb{1}_{2\times 2}$ and $t^1_{(i)}=\tau_3$, where $\tau_3$ is the
third Pauli matrix, are defined in the isospin space of the $i$th nucleon.
Isoscalar and isovector coupling constants are denoted by $c_j^0$ and $c_j^1$,
respectively. They are linearly related to the coupling constants for protons
and neutrons\footnote{This definition of $c_j^p$ and $c_j^n$ differs by a factor
of 2 with respect to the one used in, e.g., \cite{Catena:2015uha}. This is
consistent with Eq.~(\ref{eq:dsigma}) and our normalization of the nuclear
response functions. Our response functions are a factor of 4 larger than those
given in output by the Mathematica notebook
in~\cite{Anand:2013yka}.}:~$c^{p}_j=(c^{0}_j+c^{1}_j)$,
$c^{n}_j=(c^{0}_j-c^{1}_j)$, and have dimension [mass]$^{-2}$.
Equation~\eqref{eq:H_chiT} is valid under the assumption that the dark
matter--nucleus interaction is the sum of dark matter interactions with the
individual nucleons. Corrections beyond this (impulse) approximation are
discussed
in~\cite{PhysRevLett.91.231301,Cirigliano:2012pq,Menendez:2012tm,Klos:2013rwa,Hoferichter:2015ipa,PhysRevD.94.063505}.

\begin{table}[t]
    \begin{ruledtabular}
    \begin{tabular}{ll}
        $\hat{\mathcal{O}}_1 = \mathbb{1}_{\chi N}$ & $\hat{\mathcal{O}}_{9~} = i{\bf{\hat{S}}}_\chi\cdot\left({\bf{\hat{S}}}_N\times\frac{{\bf{\hat{q}}}}{m_N}\right)$  \\
        $\hat{\mathcal{O}}_3 = i{\bf{\hat{S}}}_N\cdot\left(\frac{{\bf{\hat{q}}}}{m_N}\times{\bf{\hat{v}}}^{\perp}\right)$ &   $\hat{\mathcal{O}}_{10} = i{\bf{\hat{S}}}_N\cdot\frac{{\bf{\hat{q}}}}{m_N}$   \\
        $\hat{\mathcal{O}}_4 = {\bf{\hat{S}}}_{\chi}\cdot {\bf{\hat{S}}}_{N}$ &   $\hat{\mathcal{O}}_{11} = i{\bf{\hat{S}}}_\chi\cdot\frac{{\bf{\hat{q}}}}{m_N}$   \\                                                                             
        $\hat{\mathcal{O}}_5 = i{\bf{\hat{S}}}_\chi\cdot\left(\frac{{\bf{\hat{q}}}}{m_N}\times{\bf{\hat{v}}}^{\perp}\right)$ &  $\hat{\mathcal{O}}_{12} = {\bf{\hat{S}}}_{\chi}\cdot \left({\bf{\hat{S}}}_{N} \times{\bf{\hat{v}}}^{\perp} \right)$ \\                                                                                                                 
        $\hat{\mathcal{O}}_6 = \left({\bf{\hat{S}}}_\chi\cdot\frac{{\bf{\hat{q}}}}{m_N}\right) \left({\bf{\hat{S}}}_N\cdot\frac{\hat{{\bf{q}}}}{m_N}\right)$ &  $\hat{\mathcal{O}}_{13} =i \left({\bf{\hat{S}}}_{\chi}\cdot {\bf{\hat{v}}}^{\perp}\right)\left({\bf{\hat{S}}}_{N}\cdot \frac{{\bf{\hat{q}}}}{m_N}\right)$ \\   
        $\hat{\mathcal{O}}_7 = {\bf{\hat{S}}}_{N}\cdot {\bf{\hat{v}}}^{\perp}$ &  $\hat{\mathcal{O}}_{14} = i\left({\bf{\hat{S}}}_{\chi}\cdot \frac{{\bf{\hat{q}}}}{m_N}\right)\left({\bf{\hat{S}}}_{N}\cdot {\bf{\hat{v}}}^{\perp}\right)$  \\
        $\hat{\mathcal{O}}_8 = {\bf{\hat{S}}}_{\chi}\cdot {\bf{\hat{v}}}^{\perp}$  & $\hat{\mathcal{O}}_{15} = -\left({\bf{\hat{S}}}_{\chi}\cdot \frac{{\bf{\hat{q}}}}{m_N}\right)$ \\ 
& \qquad \quad $\left[ \left({\bf{\hat{S}}}_{N} \times {\bf{\hat{v}}}^{\perp} \right) \cdot \frac{{\bf{\hat{q}}}}{m_N}\right] $                
    \end{tabular}
    \end{ruledtabular}
    \caption{Interaction operators defining the effective theory of dark
    matter--nucleon interactions. The operator $\mathbb{1}_{\chi N}$ is the
    identity in the two-particle spin space. Here $m_N$ is the nucleon mass and
    all interaction operators have the same mass dimension. For simplicity, we
    omit the nucleon index $i$.} 
    \label{tab:operators}
\end{table}

We derive the differential cross section for dark matter--nucleus scattering
from the Hamiltonian density in Eq.~(\ref{eq:H_chiT}):
\begin{equation}
\label{eq:dsigma} 
\begin{split}
\frac{{\rm d}\sigma}{{\rm d}q^2} 
&=\frac{1}{(2J+1) v^2}\sum_{\tau,\tau'} \bigg[ \\
&\sum_{\ell=M,\Sigma',\Sigma''} R^{\tau\tau'}_\ell  \left(v_{T}^{\perp 2}, {q^2 \over m_N^2} \right) W_\ell^{\tau\tau'}(q^2) \\
&+{q^{2} \over m_N^2} \sum_{m=\Phi'', \Phi'' M, \tilde{\Phi}', \Delta, \Delta \Sigma'} \hspace{-0.4 cm}R^{\tau\tau'}_m\left(v_{T}^{\perp 2}, {q^2 \over m_N^2}\right) W_m^{\tau\tau'}(q^2) \bigg] ,
\end{split}
\end{equation}
where $J$ is the target nucleus spin, $v$ is from now onwards the dark
matter--nucleus relative velocity, and $v_{q}^{\perp 2}=v^2-q^2/(4\mu_{\chi
A}^2)$. Here $q\equiv|\mathbf{q}|$ and $\mu_{\chi A}$ is the dark
matter--nucleus reduced mass. The eight dark matter response functions
$R^{\tau\tau'}_\ell$ and $R^{\tau\tau'}_m$ depend on the isoscalar and isovector
coupling constants $c_j^\tau$, $q^2/m_N^2$ and $v_{q}^{\perp 2}$. They were
first derived in~\cite{Fitzpatrick:2012ix,Anand:2013yka} and are listed in the
Appendix.

The eight nuclear response functions $W_\ell^{\tau\tau'}$ and $W_m^{\tau\tau'}$
in Eq.~(\ref{eq:dsigma}) are given by
\begin{equation}
\label{eq:W}
\begin{split}
W_{AB}^{\tau \tau^\prime}\left(q^2\right) &= \sum_{L\le 2J}  \langle J,T,M_{T} ||~ \hat{A}_{L;\tau} (q)~ || J,T,M_{T} \rangle \\
&\times\langle J,T,M_{T} ||~ \hat{B}_{L;\tau^\prime} (q)~ || J,T,M_{T} \rangle,
\end{split}
\end{equation}
where $\hat{A}_{L;\tau^\prime} (q)$ and $\hat{B}_{L;\tau^\prime} (q)$ can each
be one of the nuclear response operators defined below in Eq.~(\ref{eq:resop}).
There are six independent nuclear response functions where $A=B$ and two
interference ones with $B \neq A$. For $B=A$, we simplify the notation writing
$W_{AA}^{\tau \tau^\prime}=W_{A}^{\tau \tau^\prime}$. In Eq.~(\ref{eq:W}), $T$
and $M_{T}$ are the nuclear isospin and associated magnetic quantum number,
respectively. Matrix elements in Eq.~(\ref{eq:W}) are reduced in the spin
magnetic quantum number $M_J$ according to
\begin{equation}
\label{eq:red}
\begin{split}
\langle J,M_J |\,{M}_{LM;\tau}\,|J,M_J\rangle &=(-1)^{J-M_J}\left(
\begin{array}{ccc} J&L&J\\
-M_J&M&M_J 
\end{array} 
\right) \\
&\times \langle  J  ||\,{M}_{L;\tau}\,|| J  \rangle \,.
\end{split}
\end{equation}
The nuclear response operators in Eq.~(\ref{eq:W}) admit the following representation
\begin{eqnarray}
M_{LM;\tau}(q) &=& \sum_{i=1}^{A} M_{LM}(q \boldsymbol{\rho}_i) t^{\tau}_{(i)},\nonumber\\
\Sigma'_{LM;\tau}(q) &=& -i \sum_{i=1}^{A} \left[ \frac{1}{q} \overrightarrow{\nabla}_{\boldsymbol{\rho}_i} \times {\bf{M}}_{LL}^{M}(q \boldsymbol{\rho}_i)  \right] \cdot \vec{\sigma}_{(i)} t^{\tau}_{(i)},\nonumber\\
\Sigma''_{LM;\tau}(q) &=&\sum_{i=1}^{A} \left[ \frac{1}{q} \overrightarrow{\nabla}_{\boldsymbol{\rho}_i} M_{LM}(q \boldsymbol{\rho}_i)  \right] \cdot \vec{\sigma}_{(i)} t^{\tau}_{(i)},\nonumber\\
\Delta_{LM;\tau}(q) &=&\sum_{i=1}^{A}  {\bf{M}}_{LL}^{M}(q \boldsymbol{\rho}_i) \cdot \frac{1}{q}\overrightarrow{\nabla}_{\boldsymbol{\rho}_i} t^{\tau}_{(i)}, \nonumber\\
\tilde{\Phi}^{\prime}_{LM;\tau}(q) &=& \sum_{i=1}^A \left[ \left( {1 \over q} \overrightarrow{\nabla}_{\boldsymbol{\rho}_i} \times {\bf{M}}_{LL}^M(q \boldsymbol{\rho}_i) \right)  \hspace{-0.1 cm} \cdot \hspace{-0.1 cm} \left(\vec{\sigma}_{(i)} \times {1 \over q} \overrightarrow{\nabla}_{\boldsymbol{\rho}_i} \right) \right. \nonumber\\
&+& \left. {1 \over 2} {\bf{M}}_{LL}^M(q \boldsymbol{\rho}_i) \cdot \vec{\sigma}_{(i)} \right]~t^\tau_{(i)}, \nonumber \\
\Phi^{\prime \prime}_{LM;\tau}(q) &=& i  \sum_{i=1}^A\left( {1 \over q} \overrightarrow{\nabla}_{\boldsymbol{\rho}_i}  M_{LM}(q \boldsymbol{\rho}_i) \right) \hspace{-0.1 cm} \cdot \hspace{-0.1 cm}\left(\vec{\sigma}_{(i)} \times{1 \over q} \overrightarrow{\nabla}_{\boldsymbol{\rho}_i}  \right)t^\tau_{(i)}, \nonumber\\ 
\label{eq:resop}
\end{eqnarray}
where $\boldsymbol{\rho}_i$ is the $i$th nucleon position vector in the nucleus
center-of-mass (c.m.) frame and $\vec{\sigma}_{(i)}$ denotes the
Pauli spin matrices. In Eq.~(\ref{eq:resop}) we define $M_{LM}(q
\boldsymbol{\rho}_i)=j_{L}(q
\boldsymbol{\rho}_i)Y_{LM}(\Omega_{\boldsymbol{\rho}_i})$ and
${\bf{M}}_{LL}^{M}(q \boldsymbol{\rho}_i)=j_{L}(q \boldsymbol{\rho}_i){\bf
Y}^M_{LL1}(\Omega_{\boldsymbol{\rho}_i})$, where $\Omega_{\boldsymbol{\rho}_i}$
represents azimuthal and polar angles of $\boldsymbol{\rho}_i$; $Y_{LM}$ and
${\bf Y}^M_{LL1}$ are spherical and vector-spherical harmonics, respectively;
and $j_{L}$ are spherical Bessel functions. The nuclear response functions
$W_\ell^{\tau\tau'}$ and $W_m^{\tau\tau'}$ in Eq.~(\ref{eq:W}) depend on $q$
quadratically when single-nucleon states are expressed in the harmonic
oscillator (HO) basis.

The nuclear response operators in Eq.~(\ref{eq:W}) arise from the multipole
expansion of nuclear charges and currents produced in the scattering of dark
matter by nuclei \cite{Fitzpatrick:2012ix,Anand:2013yka}:~$M_{LM;\tau}$ arises
from the nuclear vector charge; $\Sigma'_{LM;\tau}$ and $\Sigma''_{LM;\tau}$
from the nuclear spin current; $\Delta_{LM;\tau}$ from the nuclear convection
current; and $\tilde{\Phi}'_{LM;\tau}$ and $\Phi''_{LM;\tau}$ from the nuclear
spin-velocity current. In the zero-momentum transfer limit, a simple intuitive
characterization for some of the nuclear response operators in Eq.~(\ref{eq:W})
is possible. For a given target nucleus, $M_{00;\tau}$ measures the nucleon
content, $\Sigma'_{1M;\tau}$ and $\Sigma''_{1M;\tau}$ the nucleon spin content,
$\Delta_{1M;\tau}$ the distribution of nucleon orbital angular momentum, and
$\Phi^{\prime \prime}_{00;\tau}$ the nucleon spin-orbit coupling content.

\subsection{\textit{Ab initio} nuclear response functions}
\label{sec:abinitio}
\noindent In this work we employ the \textit{ab initio} no-core shell model
(NCSM) technique \cite{navratil2009,2013PrPNP..69..131B} to evaluate the various
nuclear response functions in Eq.~\eqref{eq:W}. The starting point of NCSM
calculations is the nonrelativistic Hamiltonian for a system of $A$ nucleons
interacting by realistic nucleon--nucleon ($V_{\NN}$) and three-nucleon
($V_{\NNN}$) interactions:
\begin{equation}
\label{eq:H}
H=\sum_{i=1}^A \frac{\mathbf{\hat{p}}_i^2}{2m_N} + \sum_{i<j=1}^A \hat{V}_{\NN; ij}
+ \sum_{i<j<k=1}^A \hat{V}_{\NNN;ijk},
\end{equation}
where $\mathbf{p}_i$ are the nucleon momenta. In NCSM, the total wave function
is expanded and the Hamiltonian is diagonalized in a fully antisymmetric
many-body HO basis.

\begin{figure}[b]
\includegraphics{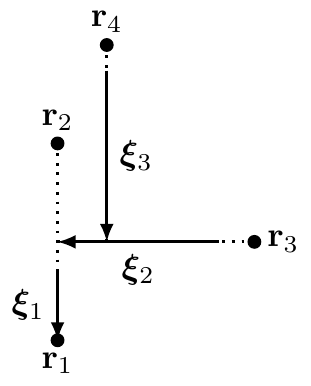}
\caption{Graphical representation of the relative Jacobi coordinates defined in
Eq.~\eqref{eq:jcoordinates} for $A=4$ nucleons.}
\label{fig:jacobi}
\end{figure}
In the present study we focus on few-body nuclear systems. In this case it is
most efficient to formulate NCSM in relative Jacobi coordinates
\cite{navratil2009}. Different sets of Jacobi coordinates can be employed, one
of which is particularly suitable for the construction of the antisymmetrized HO
basis:
\begin{equation} \label{eq:jcoordinates}
\begin{split}
\boldsymbol{\xi}_0 &= \sqrt{\frac{1}{A}} \sum_{j=1}^A \mathbf{r}_j, \\
\boldsymbol{\xi}_i &= \sqrt{\frac{i}{i+1}}\left(
\frac{1}{i}\sum_{j=1}^{i} \mathbf{r}_j - \mathbf{r}_{i+1} \right),
\end{split}
\end{equation}
with $\mathbf{r}_i$ being the coordinate of nucleon $i=1,\ldots,A$. In this set,
graphically represented in Fig.~\ref{fig:jacobi} for $A=4$ nucleons, the
coordinate $\boldsymbol{\xi}_0$ is proportional to the c.m.\ coordinate of the
$A$-body system and the coordinates $\boldsymbol{\xi}_i$, $i=1,\ldots,A-1$, are
proportional to the relative positions of nucleon $i+1$ with respect to the
c.m.\ of the $i$-nucleon subcluster. When the single-nucleon coordinates and
momenta in the Hamiltonian \eqref{eq:H} are transformed into coordinates
\eqref{eq:jcoordinates}, the kinetic term splits into a part depending only on
the c.m.\ coordinate $\boldsymbol{\xi}_0$ and an intrinsic part depending only
on the internal Jacobi coordinates $\{ \boldsymbol{\xi}_i \}_{i=1}^{A-1}$.
Translational invariance of $V_{\NN}$ and $V_{\NNN}$ interactions, i.e.\
independence of $\boldsymbol{\xi_0}$, allows us to separate out the c.m.\ term
and thus decrease the number of degrees of freedom. Consequently, the $A$-body
HO basis states can be constructed as
\begin{equation} \label{eq:pbasis}
\vert (\ldots(\alpha_1,\alpha_2)J_3T_3,\alpha_3)J_4 T_4,\ldots
,\alpha_{\Amo})J_AT_A\rangle,
\end{equation}
where $\vert \alpha_i \rangle \equiv \vert n_i(l_i s_i)j_i t_i \rangle$ are HO
states, depending on coordinates $\boldsymbol{\xi}_i$, with radial $n_i$,
orbital $l_i$, spin $s_i$, angular momentum $j_i$, and isospin $t_i$ quantum
numbers. The parentheses in \eqref{eq:pbasis} represent angular momentum and
isospin coupling. The quantum numbers $J_i$ and $T_i$ ($i=3,\ldots,A$) are
angular momentum and isospin quantum numbers of $i$-nucleon clusters. The basis
is truncated by restricting the total number of HO quanta:
\begin{equation} \label{eq:nmax}
\sum_i 2n_i+l_i \le \nm,
\end{equation}
which defines the size of the model space. NCSM calculations are thus
variational and converge to exact results with increasing $\nm$. In the case of
few-body systems, as considered in this work, calculations with sufficiently
large $\nm$ can be performed to reach satisfactory convergence.

Before the diagonalization of the Hamiltonian \eqref{eq:H} the basis states
\eqref{eq:pbasis} have to be antisymmetrized with respect to the exchanges of
all nucleons. The antisymmetrization procedure with Jacobi-coordinate HO basis
states is extensively discussed in Ref.~\cite{navratil2009} and we will only
summarize its main points here. The fully antisymmetric $A$-body HO basis is
obtained by diagonalization of the antisymmetrizer operator
$\hat{\mathcal{A}}_A$ between the basis states \eqref{eq:pbasis}. The
antisymmetrizer is defined as
\begin{equation} \label{eq:X}
\hat{\mathcal{A}}_{A} = \frac{1}{A!} \sum_{\pi} \mathrm{sgn}(\pi)\,
\hat{\mathcal{P}}_\pi,
\end{equation}
where the summation extends over all permutations $\pi$, with parity
$\mathrm{sgn}(\pi)$, of single-nucleon coordinates realized on the states
\eqref{eq:pbasis} by permutation operator $\hat{\mathcal{P}}_\pi$. The
eigenvectors of the antisymmetrizer \eqref{eq:X} span two eigenspaces --- one
corresponding to eigenvalue 1 formed by physical antisymmetric states and one
corresponding to eigenvalue 0 formed by spurious states. The antisymmetrizer
can be represented as
\begin{equation} \label{eq:factX}
\hat{\mathcal{A}}_{A} = \hat{\mathcal{A}}_{\Amo} \frac{1}{A} \left[
1-(A-1)\hat{\mathcal{P}}_{\Amo,A} \right] \hat{\mathcal{A}}_{\Amo},
\end{equation}
where the operator $\hat{\mathcal{P}}_{\Amo,A}$ interchanges the coordinates of
nucleons $A-1$ and $A$. Equation~\eqref{eq:factX} provides the basis for an iterative
procedure to obtain fully antisymmetrized states from states with a lower degree
of antisymmetry. Explicit expressions for the matrix elements of the
antisymmetrizer $\hat{\mathcal{A}}_A$ between the basis states \eqref{eq:pbasis}
can be found e.g.\ in Ref.~\cite{Navratil:1999pw}. The resulting states can be
expanded in terms of the original basis containing an antisymmetric cluster of
$A-1$ nucleons and one nucleon as
\begin{equation} \label{eq:Abasis}
\begin{split}
&\vert N_A i_A J_A T_A \rangle =\\ & \sum
\langle(N_{\Amo}i_{\Amo}J_{\Amo}T_{\Amo},\alpha_{\Amo})J_AT_A \vert N_A i_A J_A T_A \rangle\\
&\times\vert (N_{\Amo}i_{\Amo}J_{\Amo}T_{\Amo},\alpha_{\Amo})J_AT_A \rangle,
\end{split}
\end{equation}
where the expansion coefficients obtained from the eigenvectors of the
antisymmetrizer are the coefficients of fractional parentage.
Thanks to the important property of the antisymmetrizer of being diagonal in the
total number of HO quanta $N_A$ the fully antisymmetric states \eqref{eq:Abasis}
can be classified by $N_A=N_{\Amo}+2n_{\Amo}+l_{\Amo}$ ($N_2=2n_1+l_1$ for
two-nucleon states) and the quantum number $i_A$ which distinguishes different
states with the same set of quantum numbers $N_A,J_A,T_A$.

To evaluate the matrix elements of two-{} and three-body $V_{\NN}$ and
$V_{\NNN}$ potentials in the Hamiltonian \eqref{eq:H} between the
antisymmetrized many-body HO states one can recursively make use of the
expansion in Eq.~\eqref{eq:Abasis}. However, it is more efficient to employ more
suitable sets of Jacobi coordinates together with the associated HO states
containing antisymmetrized states of $A-2$ and two nucleons or $A-3$ and three
nucleons \cite{navratil2009}.

To construct the nuclear response functions defined in Eq.~\eqref{eq:W} we need
to evaluate matrix elements of the various operators in Eq.~\eqref{eq:resop}
between the ground-state wave functions of the Hamiltonian. The
nuclear matrix elements in \eqref{eq:W} can be further reduced in nuclear
isospin and written as
\begin{equation}
\begin{split}
& \langle J^\pi T M_T \vert\vert \sum_{i=1}^A \hat{A}_{L \tau}(q\boldsymbol{\rho}_i)
\vert\vert J^\pi T M_T \rangle \\
&= (-1)^{T-M_T} \begin{pmatrix} T & \tau & T \\ -M_T & 0 & M_T \end{pmatrix}
\\
&\times \langle J^\pi T \vert\vert\vert \sum_{i=1}^A \hat{A}_{L
\tau}(q\boldsymbol{\rho}_i) \vert\vert\vert J^\pi T \rangle.
\end{split}
\end{equation}
The NCSM technology for computing such nuclear matrix elements is
analogous to standard SM calculations. One-body transition
densities (OBTD) are introduced so that the many-body matrix elements
of one-body operators (reduced both in nuclear spin and isospin) can
be expressed as products of OBTD and single-particle matrix elements
\cite{Navratil:1999pw}:
\begin{equation}
\label{eq:mmet}
\begin{split}
& \langle J^\pi T \vert\vert\vert \sum_{i=1}^A \hat{A}_{L
\tau}(q\boldsymbol{\rho}_i) \vert\vert\vert  J^\pi T \rangle \\
& = \sum_{\alpha \beta} \Psi_{\alpha \beta}^{L
\tau } \langle \alpha \vert\vert\vert \hat{A}_{L
\tau} (q \boldsymbol{\rho}_A)
 \vert\vert\vert \beta \rangle,
\end{split}
\end{equation}
where $\ket{\alpha(\beta)} = \ket{ n_{\alpha(\beta)} (l_{\alpha(\beta)}
\tfrac{1}{2})j_{\alpha(\beta)} \tfrac{1}{2}}$ denote single-nucleon HO states
associated with Jacobi coordinate $\boldsymbol{\xi}_{\Amo} =
-\sqrt{A/(A-1)}\boldsymbol{\rho}_A$.
The OBTD $\Psi_{\alpha \beta}^{L \tau}$ is given by
\begin{align} \label{eq:obtd}
\begin{split}
\Psi_{\alpha\beta}^{L \tau}
&= A \sum 
\langle J^\pi T \vert (N_{\Amo} i_{\Amo}J_{\Amo}T_{\Amo}, \alpha )JT \rangle \\
&\times \langle (N_{\Amo}i_{\Amo}J_{\Amo}T_{\Amo}, \beta )JT \vert J^\pi T \rangle \\
&\times \widehat{J}^2(-1)^{J_{\Amo}+L + J +j_\beta } \begin{Bmatrix}J_{\Amo} & j_\beta & J \\ L & J & j_\alpha\end{Bmatrix} \\
&\times \widehat{T}^2(-1)^{T_{\Amo}+\tau + T +\frac{1}{2} }
\begin{Bmatrix}T_{\Amo} & \frac{1}{2} & T \\ \tau & T &
\frac{1}{2}\end{Bmatrix},
\end{split}
\end{align}
where the terms in curly brackets are the Wigner $6j$ symbols and we used expansion of
the eigenstate in the basis \eqref{eq:Abasis}. In Eq.~\eqref{eq:mmet}, the
single-particle matrix elements reduced in both angular momentum and isospin can
be simplified by using \( \langle \frac{1}{2} \vert\vert t^\tau \vert\vert
\frac{1}{2} \rangle = \sqrt{2(2\tau +1)}\):
\begin{equation}
\langle \alpha \vert\vert\vert \hat{O}_{L \tau}(q\boldsymbol{\rho}_A)
 \vert\vert\vert\beta \rangle = \sqrt{2(2\tau +1)}
\langle \alpha \vert\vert \hat{O}_{L}(q\boldsymbol{\rho}_A) \vert\vert \beta \rangle,
\end{equation}
where \(\langle \alpha \vert\vert \hat{O}_{L} (q\boldsymbol{\rho}_A) \vert\vert\vert \beta
\rangle\) are single-particle matrix elements reduced in angular momentum
only. In a HO basis these matrix elements can be calculated
analytically and are listed e.g.\ in Ref.~\cite{Fitzpatrick:2012ix}.

\subsection{Chiral nuclear interactions}
\label{sec:chiral}
\noindent The theory of nuclear forces has a long history---starting with the
seminal meson-exchange hypothesis of Yukawa. The current state of the art
involves the use of chiral EFT and has opened the door for a description of
atomic nuclei consistent with the underlying symmetries of QCD. Nuclear
interactions from chiral EFT are based on the use of nucleons and pions as the
relevant degrees of freedom, but employ symmetries and the pattern of
spontaneous symmetry breaking of
QCD~\cite{vanKolck:1994yi,Epelbaum:2008ga,Machleidt:2011zz}. In this approach,
the exchange of pions within chiral perturbation theory yields the long-ranged
contributions of the nuclear interaction, while short-ranged components are
included as contact terms. Regularization is needed to deal with divergent
momentum-space integrals. The interaction is parametrized in terms of low-energy
constants (LECs) that, in principle, can be connected to QCD predictions.
However, the currently viable approach to accurately describe atomic nuclei in
chiral EFT requires that the LECs are constrained from experimental low-energy
data. The bulk of this fit data traditionally consists of cross sections
measured in nucleon--nucleon scattering experiments. The interactions from
chiral EFT exhibit a power counting in the ratio $Q/\Lambda$, with $Q$ being the
low-momentum scale that is characteristic for the nuclear observable under
consideration and $\Lambda$ the EFT breakdown scale, which is of the order of
1~GeV. In this approach, three-nucleon forces enter at next-to- next-to-leading
order (NNLO). Both regulator independence and an accurate power counting scheme
are crucial ingredients for the EFT approach to nuclear forces. In combination
these properties allow for order-by-order improvement with decreasing truncation
error, where the magnitude of such errors can also be quantified.

In this work, the nuclear interaction enters in the many-body
Hamiltonian~\eqref{eq:H} that is diagonalized in a basis to yield the nuclear
wave function. In order to capitalize on recent developments in the
quantification of uncertainties of nuclear
forces~\cite{Ekstrom:2013hd,2015JPhG...42c4003E,NavarroPerez:2015fb,Carlsson:2016cw}
we employ the family of 42 different interactions at NNLO (labeled NNLO$_{\rm
sim}$) that was constructed by \textcite{Carlsson:2016cw}. These potentials are
optimized to simultaneously reproduce $\NN$ as well as $\pi N$-scattering data,
the binding energies and charge radii of \nuc{2,3}{H} and \nuc{3}{He}, the
quadrupole moment of \nuc{2}{H}, as well as the $\beta$-decay half-life of
\nuc{3}{H}. Utilizing such a large set of interactions allows us to better
explore the systematic uncertainties. Each NNLO$_{\rm sim}$ potential is
associated with one of seven different regulator cutoffs $\Lambda_{\rm EFT} =
450, 475, \ldots, 575, 600$ MeV. In addition, the database of experimental $\NN$
scattering cross sections used to constrain the respective interaction was also
varied. It was truncated at six different maximum scattering energies in the
laboratory system $T_{\rm Lab}^{\rm max} = 125, \ldots, 290$ MeV. It should be
pointed out that for all NNLO$_{\rm sim}$ interactions an equally good
description of the fit data is attained and that all LECs are of natural size.
See Ref.~\cite{Carlsson:2016cw} for a complete description. In this work we are
mainly interested in the nuclear wave functions of \nuc{3,4}{He}. We note that
the binding energy of \nuc{3}{He} is accurately described for all these
interactions since it is included in the pool of fit data. Predictions for
$E(\nuc{4}{He})$ vary within a $\sim 2$~MeV range around the experimental
binding energy.

\section{Results}
\label{sec:results}
\noindent The main focus of this work is to quantify the impact of systematic
nuclear structure uncertainties on the interpretation of data from dark matter
searches. In the present study we consider only light nuclear systems that can
be calculated reliably and accurately without uncontrollable approximations. In
particular, we performed \textit{ab initio} NCSM calculations of \nuc{3}{He} and
\nuc{4}{He} and constructed all relevant nuclear response functions that appear
in elastic dark matter--nucleus scattering. The generated response functions
were then employed to explore the sensitivity of selected physical observables
to nuclear-structure uncertainties.

\subsection{Nuclear response functions of \nuc{3}{He} and \nuc{4}{He}}
\label{sec:W}
\noindent In this section we present the nuclear response functions defined in
Eq.~\eqref{eq:W}. In order to evaluate the nuclear matrix elements in
Eq.~\eqref{eq:mmet} and construct the nuclear response functions we performed
\textit{ab initio} NCSM calculations of \nuc{3}{He} and \nuc{4}{He} ground-state
wave functions using the complete family of all 42 NNLO${}_{\text{sim}}$ chiral
nuclear Hamiltonians. The NCSM model spaces used in these calculations are very
large, $\nm=40(20)$ for \nuc{3}{He}(\nuc{4}{He}), so that both energies and wave
functions are fully converged. The systematic uncertainties in the determination
of the underlying $\NN$ and $\NNN$ interactions are probed through the use of a
large family of interactions. These uncertainties propagate into the set of
calculated nuclear wave functions and thus manifest themselves as uncertainties
in the determination of the nuclear response functions. The types of the nuclear
responses generated by a particular nucleus depend on the total nuclear
ground-state angular momentum and isospin as well as on the details of the
nuclear structure.

\begin{figure}[t]
\includegraphics[height=0.32\textheight]{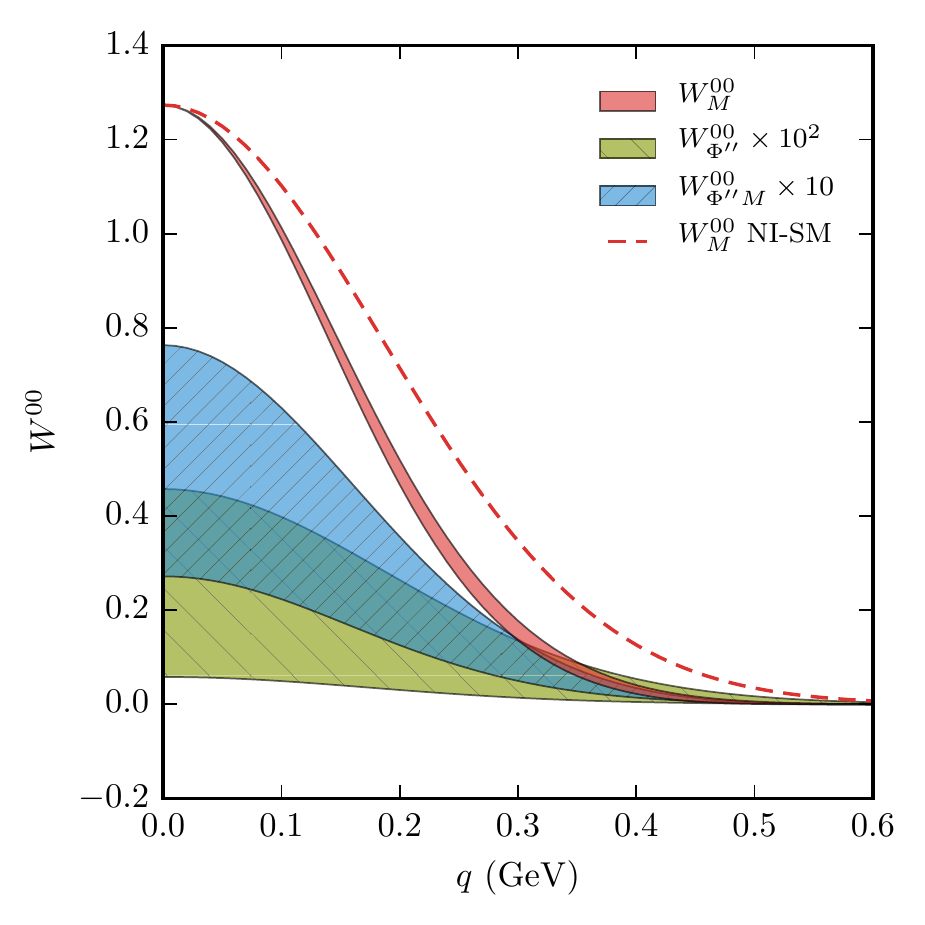}
\caption{Isoscalar nuclear response functions $W_M^{00}$, $W_{\Phi''}^{00}$
and $W_{\Phi''M}^{00}$ of \nuc{4}{He} as functions of the transferred momentum $q$,
calculated within \textit{ab initio} NCSM (shaded regions) and the
noninteracting shell model (NI-SM) (dashed
line).}
\label{fig:4He_W00all}
\end{figure}
In the case of \nuc{4}{He} most of the response functions are identically zero
due to the $J=0$ and $T=0$ ground-state quantum numbers. The only nonvanishing
nuclear response functions are the isoscalar spin-independent $W^{00}_{M}$,
$W^{00}_{\Phi''}$ and $W^{00}_{\Phi''M}$ responses, which are shown in
Fig.~\ref{fig:4He_W00all} as functions of the transferred (recoil) momentum $q$.
All NNLO${}_\text{sim}$ chiral nuclear Hamiltonians were used to calculate a
\nuc{4}{He} ground-state wave function and to evaluate the response functions.
These different curves turn out to be evenly distributed in regions that are
then represented by shaded bands in the figures. In that way, the response
functions calculated with the \textit{ab initio} NCSM technique reflect the
systematic uncertainty in the underlying nuclear interaction. The
nuclear-structure uncertainties affect the various response functions very
differently. While the dominant nuclear response $W_M^{00}$ is determined fairly
accurately, the $W^{00}_{\Phi''}$ and $W^{00}_{\Phi''M}$ responses suffer from
large uncertainties, in particular in the region of low recoil momenta $q
\approx 0$~GeV. It is to be noted that $W^{00}_{\Phi''}$ and $W^{00}_{\Phi''M}$
appear suppressed by a factor of $q^2/m_N^2$ in the scattering cross section
\eqref{eq:dsigma} and the large uncertainties are thus suppressed in the
physical observables, as will be demonstrated in Sec.~\ref{sec:dmpheno}.
The large uncertainties found in the $W^{00}_{\Phi''}$ (and consequently
$W^{00}_{\Phi''M}$) response compared to $W^{00}_{M}$ can be understood by
examining the long-wavelength limit ($q\rightarrow 0$) of the leading multipoles
of the corresponding nuclear operators~\cite{Fitzpatrick:2012ix}. In
this limit we have
\begin{equation} \label{eq:longW_M}
M_{00;0}(q) \xrightarrow{q \to 0}
\frac{1}{\sqrt{4\pi}} 
\sum_{i=1}^A \mathbb{1}_{(i)},
\end{equation}
which implies that the $W_M^{00}(q \to 0)$ response is proportional to $A^2$
independent of the nuclear dynamics. On the other hand, since
\begin{equation} \label{eq:longW_Phipp}
\Phi''_{00;0}(q) \xrightarrow{q \to 0}
\frac{-1}{3\sqrt{4\pi}}
\sum_{i=1}^A \boldsymbol{\sigma}_{(i)}\cdot\mathbf{l}_{(i)},
\end{equation}
the $W_{\Phi''}^{00}(q\rightarrow 0)$ response is proportional to the square of
the expectation value of the nucleon spin-orbit coupling in the nuclear ground
state. This quantity is difficult to access experimentally and its value is
therefore not constrained. Consequently, the $q\rightarrow 0$ behavior of the
$W^{00}_{\Phi''}$ and $W^{00}_{\Phi''M}$ responses are predictions of the
nuclear model, clearly very sensitive to the underlying nuclear Hamiltonian.

Furthermore, the functional dependence of the nuclear response
functions on the recoil momentum $q$ can be understood from
expressions \eqref{eq:W} and \eqref{eq:mmet}. In a HO basis the
single-particle matrix elements in Eq.~\eqref{eq:mmet} can be
evaluated analytically, yielding expression of the form
$P(y)\,\mathrm{e}^{-y}$, where $P(y)$ is a
polynomial~\cite{Fitzpatrick:2012ix} and $y=(qb/2)^2$ a dimensionless
quantity with $b=\sqrt{\hbar/(m_N\omega)}$ the HO length.
Since the one-body transition densities are independent of $q$, the nuclear
response functions follow this exponential suppression and their absolute
uncertainties decrease with $q$. On the other hand, the relative uncertainties
in the response functions increase for larger values of recoil momentum.
\begin{figure}[t]
\includegraphics[height=0.32\textheight]{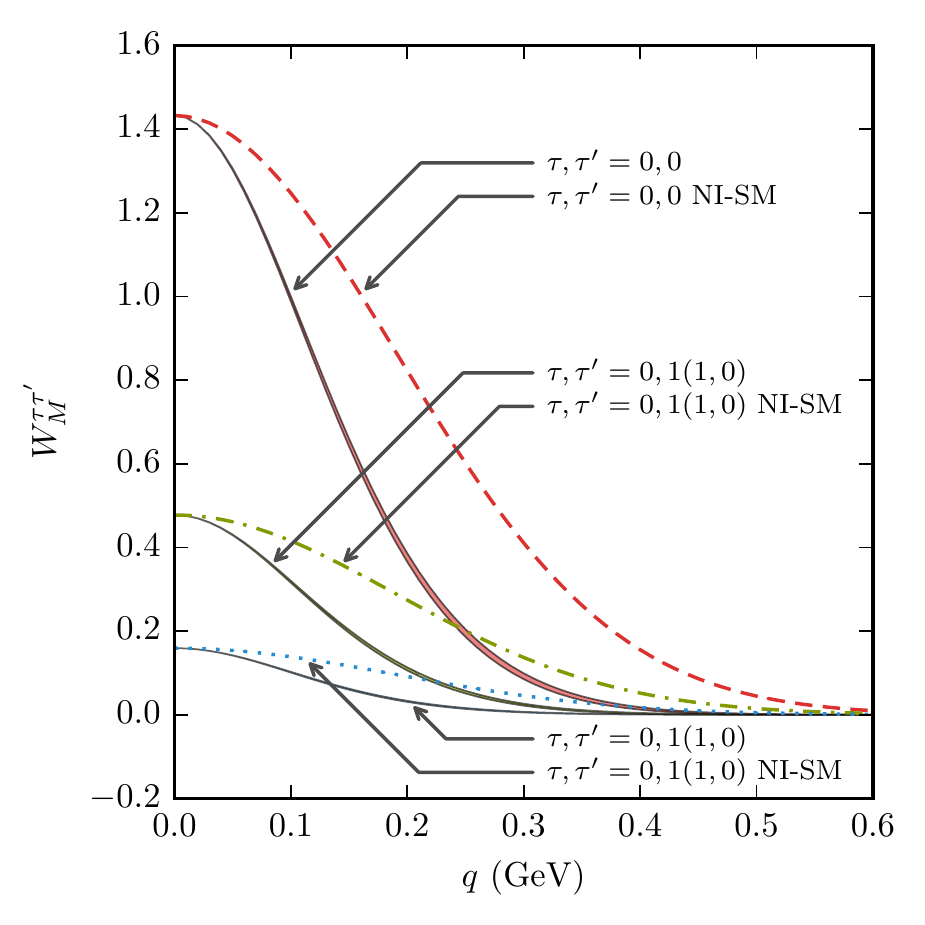}
\caption{Nuclear response functions $W_M^{\tau\tau'}$ of \nuc{3}{He} as
functions of the transferred momentum $q$, calculated within the \textit{ab
initio} NCSM (shaded regions) and the NI-SM (dashed, dashed-dotted an dotted lines).}
\label{fig:3He_WM}
\end{figure}
For comparison, also shown in Fig.~\ref{fig:4He_W00all} is the $W_M^{00}$
response function taken from Ref.~\cite{Catena:2015uha} in which SM calculations
were performed for a number of light elements in the context of dark
matter--nucleus scattering. In the particular cases of \nuc{3,4}{He}, the
interaction that was used in that work did not allow coupling between the
$0s_{1/2}$ shell and higher-lying orbits. As a consequence, the ground-state
configuration is a single Slater determinant with all four nucleons in the
$0s_{1/2}$ shell implying that the results correspond to a noninteracting shell
model (NI-SM). Even though this situation is an extreme limiting case for the SM
it is worth pointing out the main differences and new features of \textit{ab
initio} calculations. In general, the $W^{00}_{M}$ response functions generated
within the SM approach using phenomenological interactions and the NCSM approach
using chiral EFT interactions will match for recoil momentum $q=0$~GeV, due to
the $A^2$-normalization, and for large values of $q$, where the response must
vanish. Except for these limits, the results will differ. In particular, for
\nuc{4}{He} the NI-SM give larger values of $W^{00}_M$ compared to NCSM
calculations. Moreover, it is to be stressed that $W^{00}_M$ is the only
nonzero nuclear response function resulting from the NI-SM calculation, as it
includes only one single-particle orbital. In general, SM calculations with
residual valence space interactions will employ a very restricted
single-particle basis. The calculation of the nuclear response should therefore
be made with operators that have been properly renormalized to act only within
this truncated model space. In contrast, the nuclear responses in
Fig.~\ref{fig:4He_W00all} calculated within the NCSM methodology were obtained
employing a substantially larger model space. The NCSM method allows to
systematically increase the size of the model space. For these calculations a
truncation of $\nm = 20$ was used to reach full convergence with the use of bare
operators. The NCSM model space is able to accommodate details of the nuclear
structure that are crucial to expose the full complexity of the nuclear
response. This difference becomes even more evident for the $W_{\Phi''}^{00}$
response, which is evaluated as zero in the restricted $\nm=0$ ($0s_{1/2}$)
model space but is non-zero in the NCSM as it receives contributions from
nucleons that occupy higher orbitals.

\begin{figure*}[t]
\includegraphics[width=0.9\textwidth]{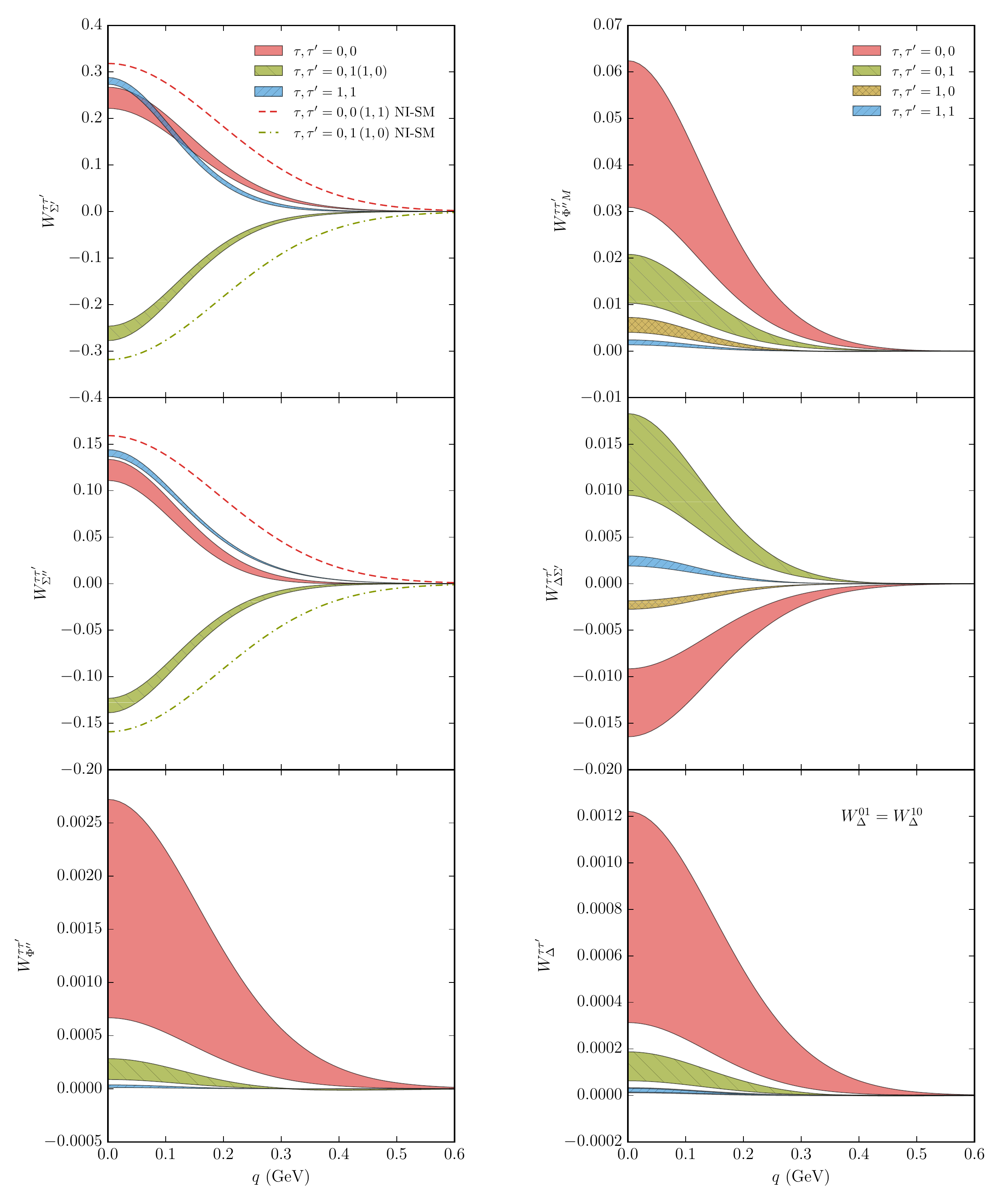}
\caption{Nuclear response functions of \nuc{3}{He} as functions of the
transferred momentum $q$ calculated within \textit{ab initio} NCSM (shaded
regions) and NI-SM (dashed and dashed-dotted lines).}
\label{fig:3He_therest}
\end{figure*}

Similar conclusions hold for the nuclear response functions of \nuc{3}{He} as
shown in Figs.~\ref{fig:3He_WM} and \ref{fig:3He_therest}. In this case there
are more nonvanishing response functions due to the $J=\tfrac{1}{2}$ and
$T=\tfrac{1}{2}$ ground-state quantum numbers. In particular, among all the
response functions in Eq.~\eqref{eq:W} only the $W_{\tilde{\Phi}'}^{\tau\tau'}$
response vanishes, since it contributes for nuclei with total angular momentum
$J \ge 1$. The dominant nuclear responses of \nuc{3}{He}, resulting both from
\textit{ab initio} NCSM calculations and the NI-SM~\cite{Catena:2015uha}, are
the spin-independent responses $W^{\tau\tau'}_{M}$, shown in
Fig.~\ref{fig:3He_WM}, and the spin-dependent response functions
$W^{\tau\tau'}_{\Sigma'}$ and $W^{\tau\tau'}_{\Sigma''}$, shown in the left
panel of Fig.~\ref{fig:3He_therest}. The nuclear structure uncertainties in the
determination of the $W^{\tau\tau'}_{M}$ response are negligibly small, making
the corresponding bands in Fig.~\ref{fig:3He_WM} almost invisible. As in the
case of \nuc{4}{He}, the \textit{ab initio} NCSM $W^{\tau\tau'}_{M}$ response
functions are smaller than the ones from the NI-SM over the whole range of
relevant recoil momenta, except for $q \rightarrow 0$ where they must agree
due to the $A^2$-normalization. The spin-dependent responses
$W^{\tau\tau'}_{\Sigma'}$ and $W^{\tau\tau'}_{\Sigma''}$ are generated by the
$\Sigma_{LM;\tau}'$ and $\Sigma_{LM;\tau}''$ nuclear operators whose leading
multipoles are proportional to the total nuclear spin operator in the
$q\rightarrow 0$ limit~\cite{Fitzpatrick:2012ix}. Similarly as for the
$\Phi_{LM;\tau}''$ operator, the ground-state expectation value of the nuclear
spin, $\frac{1}{2}\sum_{i=1}^A \vec{\sigma}_{(i)}$, is not imposed as a strict
constraint and its value can vary for different nuclear Hamiltonians. However,
the nuclear uncertainties affect these response functions only moderately. The
\textit{ab initio} NCSM calculations generate additional nuclear responses not
appearing in the NI-SM, namely the $W^{\tau\tau'}_{\Phi''}$ and
$W^{\tau\tau'}_{\Delta}$ response functions. These, in turn, generate the
interference responses $W^{\tau\tau'}_{\Phi''M}$ and $W^{\tau\tau'}_{\Delta
\Sigma'}$. Finally, the leading multipole of the nuclear response operator
$\Delta_{LM;\tau}$ is proportional to the total nuclear angular momentum,
$\sum_{i=1}^A \vec{l}_{(i)}$ \cite{Fitzpatrick:2012ix} and its expectation value
is also not imposed as a constraint on the nuclear Hamiltonian. Consequently, as
shown in Fig.~\ref{fig:3He_therest}, all these response functions exhibit large
systematic uncertainties that are, however, suppressed in physical observables
by a factor of $q^2/m_N^2$. Furthermore, isovector responses are generally
smaller in magnitude since they result from a proton-neutron difference rather
than a sum of proton and neutron contributions.

\subsection{Impact on dark matter searches}
\label{sec:dmpheno}
\begin{figure}
\includegraphics[height=0.32\textheight]{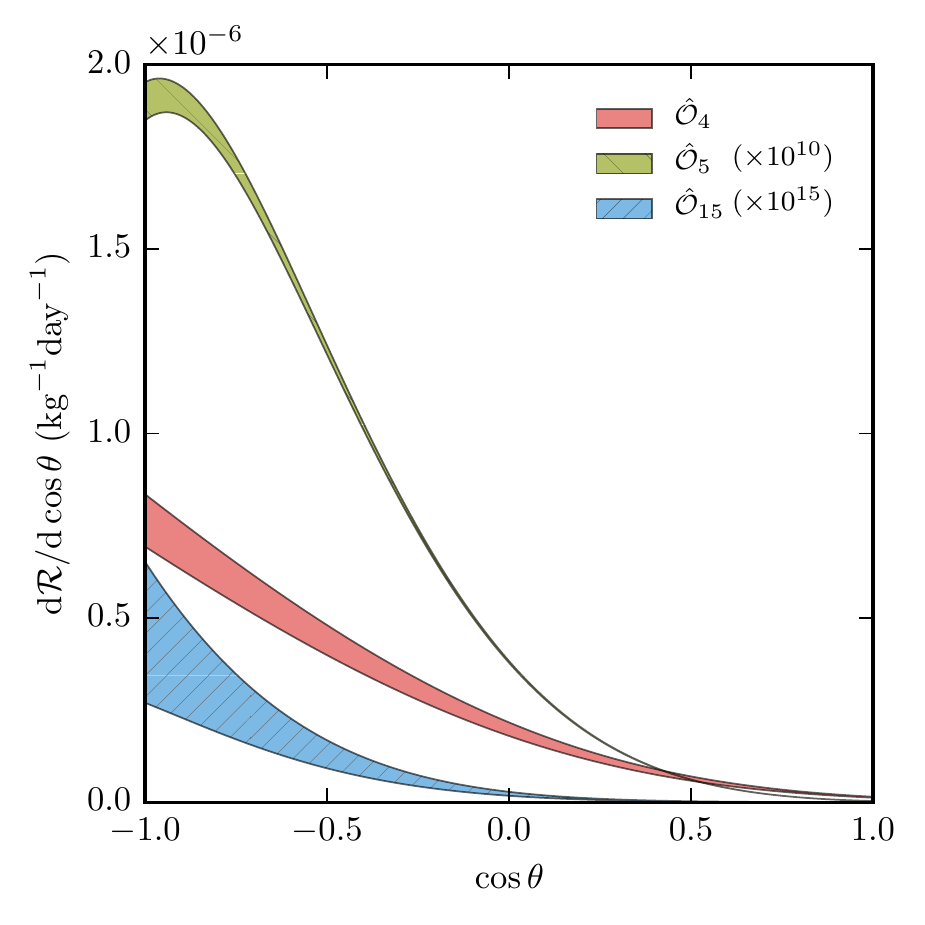}
\caption{Differential rate of nuclear recoil events as a function of the
recoil direction. We focus on the isoscalar component of the interaction
operators listed in the legend and consider \nuc{3}{He} as a target material. We
set $m_\chi=10$~GeV and the isoscalar coupling constants to the reference value
as detailed in the text. For the nuclear response functions needed in this
calculation, Eq.~(\ref{eq:W}), we consider two distinct choices, corresponding
to the lower and upper boundaries of the associated nuclear physics uncertainty
band found in Sec.~\ref{sec:abinitio}. This produces the colored bands in the
figure, which therefore account for the nuclear physics uncertainties in the
recoil rates.}
\label{fig:He3_isos}
\end{figure}
\noindent Nuclear physics uncertainties in the response functions affect the
interpretation of data from dark matter search experiments. In this subsection
we quantitatively address this matter in the context of directional dark matter
detection for which detectors with \nuc{3,4}{He} target materials are currently
in a research and development stage. The aim is to assess the impact of nuclear
physics uncertainties for these isotopes on physical observables.

We start by reviewing the basic concepts of directional dark matter detection.
The Earth's motion in the galactic rest frame induces a flux of dark matter
particles across the surface of the planet. If dark matter interacts with
nuclei, low-background experiments might be able to detect nuclear recoils
induced by the scattering of dark matter particles in a target
material~\cite{Goodman:1984dc}. The angular distribution of such nuclear recoil
events is expected to be anisotropic, as the Earth's motion in the galactic rest
frame selects a preferred direction in the sphere of recoil
directions~\cite{Spergel:1987kx}. Depending on the interaction operator in
analysis, recoil events are mainly expected in the direction opposite to the
observer's motion, or in a ring around
it~\cite{Catena:2015vpa,Kavanagh:2015jma}. In order to exploit this information,
directional dark matter detectors have been designed. They search for
anisotropies in the distribution of nuclear recoil events in low-background
underground experiments. Here we consider hypothetical detectors made of
\nuc{3}{He} or \nuc{4}{He}.

Let us now focus on physical observables. The double differential rate of
nuclear recoil events per unit detector mass is given by
\begin{equation}
\label{eq:d2R}
\begin{split}
\frac{{\rm d}^2\mathcal{R}}{{\rm d}q^2\,{\rm d}\Omega} &=  \alpha 
 \int {\rm d}^3{\bf v} \, \delta({\bf v}\cdot {\bf w}-w_{q})\, f({\bf v} + {\bf
 v_\oplus}(t))\\
 &\times v^2 \frac{{\rm d}\sigma(v^2,q^2)}{{\rm d}q^2},
\end{split}
\end{equation}
where $\alpha=\rho_\chi/(2\pi\,m_\chi m_{A})$,
$\rho_\chi=0.4$~GeV~cm$^{-3}$~\cite{Catena:2009mf} is the local dark matter
density, $m_{A}$ the target nucleus mass, ${\bf w}$ a unit vector pointing in
the nuclear recoil direction, $w_{q}=q/(2\mu_{\chi A})$ the minimum velocity
required to transfer a momentum $q$ in the scattering, and ${\bf v_\oplus}(t)$
the time-dependent Earth's velocity in the galactic rest frame. From now
onwards, we assume azimuthal symmetry around the direction of ${\bf
v_\oplus}(t)$, i.e. ${\rm d}\Omega=2\pi {\rm d}\hspace{-0.5mm}\cos\theta$, and
measure the angle $\theta$ with respect to ${\bf v_\oplus}(t)$. We approximate
the velocity distribution $f({\bf v} + {\bf v_\oplus}(t))$ with a Gaussian
function truncated at an escape velocity of 533~km~s$^{-1}$, and assume a local
standard of rest of 220~km~s$^{-1}$~\cite{Catena:2011kv,Bozorgnia:2013pua}. The
velocity integral in Eq.~(\ref{eq:d2R}) is a Radon transform. In the Gaussian
approximation, it has been evaluated analytically for all operators in
Table~\ref{tab:operators} in Ref.~\cite{Catena:2015vpa}. The key physical
observable for the present analysis is the differential rate of nuclear recoil
events per unit detector mass. This can be calculated from Eq.~(\ref{eq:d2R}) as
follows:
\begin{equation}
\frac{{\rm d}\mathcal{R}}{{\rm d}\hspace{-0.5mm}\cos\theta} = 2\pi\int_{q^2>q^2_{\rm th}} \,\frac{{\rm d}^2\mathcal{R}}{{\rm d}q^2\,{\rm d}\Omega} \,{\rm d}q^2 \,,
\label{eq:dR}
\end{equation}
where $E_{\rm th}\equiv q^2_{\rm th}/(2 m_{A})$ is the detector energy
threshold. Here we set $E_{\rm th}=0$, and assume infinite energy and angular
resolution.

\begin{figure}
\includegraphics[height=0.32\textheight]{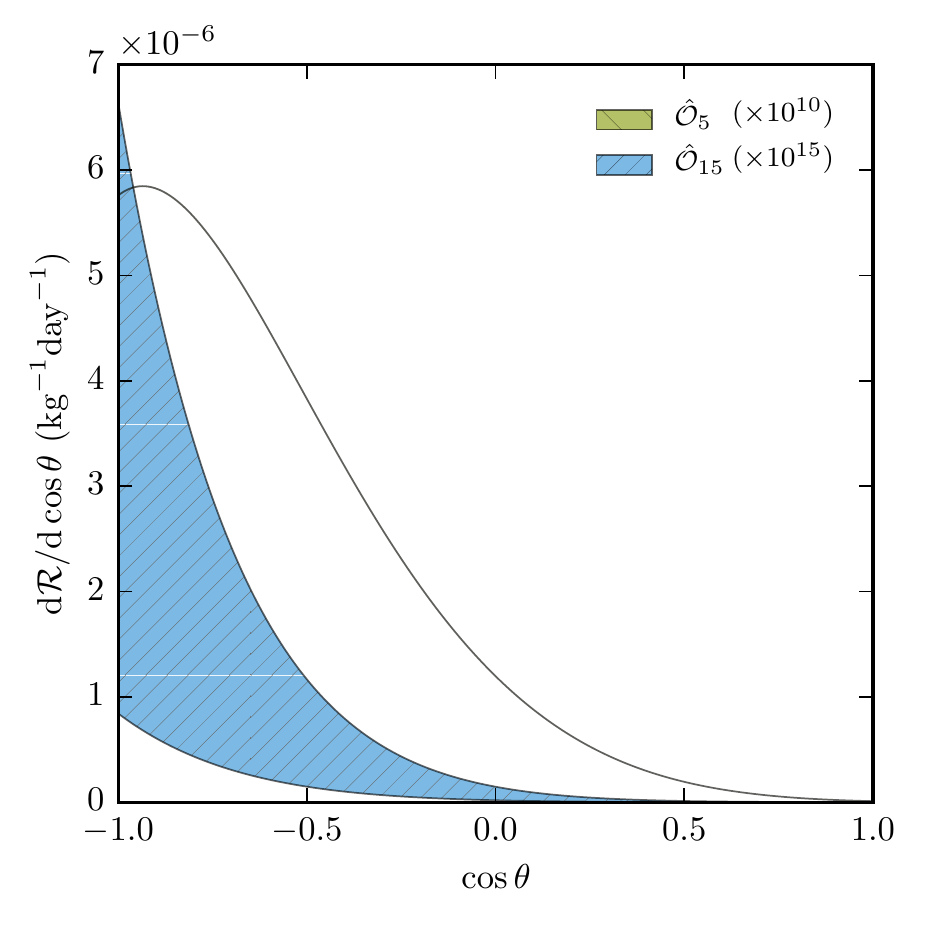}
\caption{Same as for Fig.~\ref{fig:He3_isos}, but now for a \nuc{4}{He}
  detector. Here we do not consider the interaction operator
  $\hat{\mathcal{O}}_4$ since the nuclear response functions
  $W_{\Sigma'}^{\tau\tau'}$ and $W_{\Sigma''}^{\tau\tau'}$ are
  identically zero for \nuc{4}{He}.\label{fig:He4_isos}}
\end{figure}

We now evaluate Eq.~(\ref{eq:dR}) for selected interaction operators,
namely, $\hat{\mathcal{O}}_{4}$, $\hat{\mathcal{O}}_{5}$,
$\hat{\mathcal{O}}_{9}$, $\hat{\mathcal{O}}_{10}$, $\hat{\mathcal{O}}_{12}$ and
$\hat{\mathcal{O}}_{15}$. The operator $\hat{\mathcal{O}}_{4}$ is the standard
spin-dependent interaction. It arises as the leading interaction operator from
the nonrelativistic reduction of renormalizable Lagrangians for spin 1/2 or 1
dark matter interacting with nucleons through the exchange of a heavy spin-1
particle. The operator $\hat{\mathcal{O}}_5$ can only be generated as the leading
interaction operator if dark matter has spin 1 and interacts with nucleons
through the exchange of a heavy spin-1 particle. In contrast, the operator
$\hat{\mathcal{O}}_{10}$ can arise for all dark matter particle spins, including
spin 0. Finally, the operator $\hat{\mathcal{O}}_{12}$ is always generated in
association with the operator $\hat{\mathcal{O}}_{1}$. We refer
to~\cite{Dent:2015zpa} for a comprehensive list of scenarios. Some of the
considerations above might be affected by operator
evolution~\cite{Crivellin:2014qxa,DEramo:2016gos}. 

In the following we will focus on the contribution of the isoscalar ($\tau=0$)
or isovector ($\tau=1$) component of a single operator $\hat{\mathcal{O}}_j
t^\tau$ at a time by setting only the corresponding coupling constant $c_j^\tau$
different from zero. In this case the value of the coupling constant is
$c_j^\tau=10^{-3}/m_V^2$, where $m_V=246.2$~GeV is the electroweak scale. The
value $10^{-3}/m_V^2$ is arbitrary and corresponds to the reference
WIMP--nucleon cross section $(\mu_{\chi N}^2/m_V^4)/(4\pi) \sim 7\times
10^{-45}$~cm$^2$ at $m_\chi=50$~GeV, with $\mu_{\chi N}$ the WIMP--nucleon
reduced mass. Since the rate depends quadratically on the coupling constants,
the results can be easily rescaled to other values of $c_j^\tau$.

Figure~\ref{fig:He3_isos} shows the differential rate of nuclear recoil events,
Eq.~(\ref{eq:dR}), as a function of the recoil direction $\cos\theta$. In the
figure we focus on the isoscalar component of selected interaction operators,
and consider \nuc{3}{He} as a target material. The dark matter particle mass has
been set to $m_\chi=10$~GeV and the coupling constants of the three operators in
the legend to the reference value as specified above. For the nuclear response
functions needed in this calculation, Eq.~(\ref{eq:W}), we consider two distinct
choices, corresponding to lower and upper boundary of the associated nuclear
physics uncertainty band found in Sec.~\ref{sec:W}. This produces the colored
bands reported in the figure. They describe the impact of nuclear physics
uncertainties on the physical observable considered in this investigation.
Specifically, in Fig.~\ref{fig:He3_isos} we consider the following
interactions:~the operator $\hat{\mathcal{O}}_4$, which generates the
$W_{\Sigma'}^{\tau\tau'}$ and $W_{\Sigma''}^{\tau\tau'}$ responses; the operator
$\hat{\mathcal{O}}_5$, which generates $W_{M}^{\tau\tau'}$ as the leading response;
and, finally, the operator $\hat{\mathcal{O}}_{15}$, which as the leading response
generates $W_{\Phi''}^{\tau\tau'}$. In agreement with Sec.~\ref{sec:W}, nuclear
physics uncertainties are large for $\hat{\mathcal{O}}_{15}$, moderate for
$\hat{\mathcal{O}}_{4}$, and small for $\hat{\mathcal{O}}_{5}$. For a given
interaction operator, the leading response function can be determined from
Eq.~(\ref{eq:R}) and the results in Sec.~\ref{sec:W}.

\begin{figure}
\includegraphics[height=0.32\textheight]{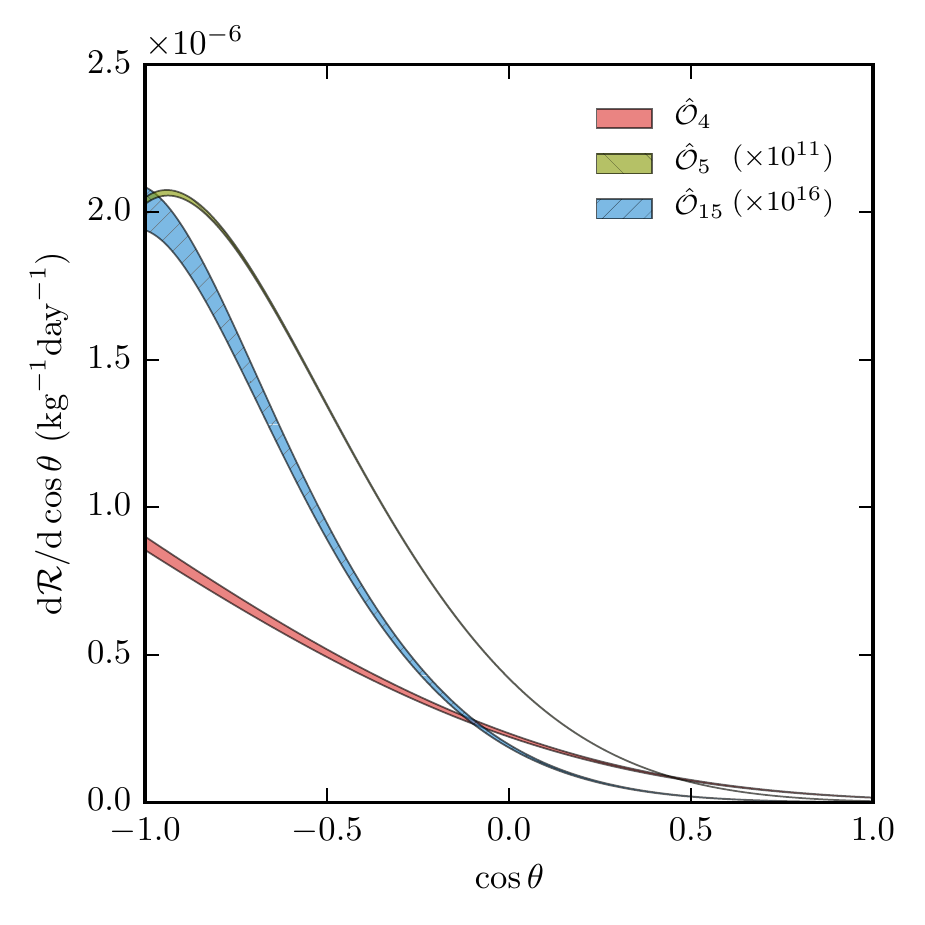}
\caption{Same as for Fig.~\ref{fig:He3_isos}, but for the isovector
   component of the operators in the legend.\label{fig:He3_isov}}
\end{figure}

\begin{figure}
\includegraphics[height=0.32\textheight]{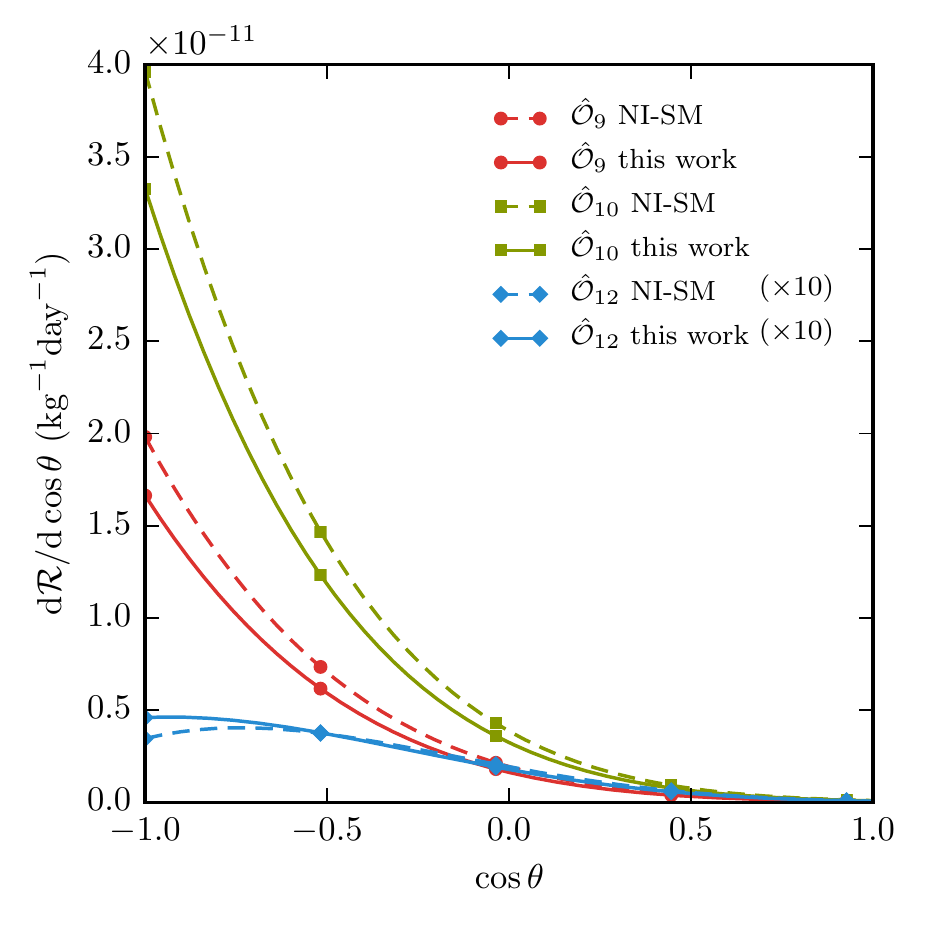}
\caption{Differential rate of nuclear recoil events computed within the {\it
ab initio} NCSM approach (this work) compared with the NI-SM rate found
in~\cite{Catena:2015vpa}. We focus on the isoscalar component of the interaction
operators listed in the legend and consider \nuc{3}{He} as a target material. We
set $m_\chi=10$~GeV and the coupling constants to the reference value given in
the text. The comparison is performed considering for each {\it ab initio}
nuclear response function the upper boundary of the corresponding nuclear
physics uncertainty band found in Sec.~\ref{sec:W}.}
\label{fig:He3_shell}
\end{figure}

Figure~\ref{fig:He3_isos} also shows that nuclear physics uncertainties are more
pronounced at $\cos\theta=-1$, where nuclear recoil rates are large. Around this
direction the integral in Eq.~(\ref{eq:dR}) is dominated by small values of $q$,
and in the $q\rightarrow 0$ limit uncertainties in the response functions grow.
We conclude that physical observables are particularly sensitive to the large
uncertainties we have found in the $q\rightarrow 0$ limit of some of the nuclear
response functions in Eq.~(\ref{eq:W}). This is one of the key results of the
present analysis.

Figure~\ref{fig:He4_isos} shows the differential rate of nuclear recoil events as
a function of $\cos\theta$ for a \nuc{4}{He} detector and the interaction
operators $\hat{\mathcal{O}}_{5}$ and $\hat{\mathcal{O}}_{15}$. We do not
consider the interaction operator $\hat{\mathcal{O}}_4$, since the nuclear
response functions $W_{\Sigma'}^{\tau\tau'}$ and $W_{\Sigma''}^{\tau\tau'}$ are
identically zero for \nuc{4}{He}. Dark matter particle mass and coupling
constants are set as above. As in the case of \nuc{3}{He}, we find that nuclear
physics uncertainties are large for $\hat{\mathcal{O}}_{15}$, which generates
the $W_{\Phi''}^{\tau\tau'}$ response. They are negligible for
$\hat{\mathcal{O}}_{5}$, which generates the $W_{M}^{\tau\tau'}$ response.

Figure~\ref{fig:He3_isov} illustrates results analogous to those reported in
Fig.~\ref{fig:He3_isos}, but now for the isovector component of the operators in
the legend. For isovector dark matter--nucleon couplings, nuclear physics
uncertainties are only moderate. From the experimental perspective, we therefore
conclude that this is the most favorable particle physics scenario.

In Fig.~\ref{fig:He3_shell} we compare two independent calculations of the
differential rate of nuclear recoil events, Eq.~(\ref{eq:dR}). The first
calculation is the one we perform here in the {\it ab initio} NCSM approach; the
second one has been performed in~\cite{Catena:2015vpa} using the nuclear SM
technique. Results are presented for the isoscalar components of selected
interaction operators, setting $m_\chi$ and associated coupling constants as
above, and focusing on \nuc{3}{He} as a target material. The comparison is
performed considering for each {\it ab initio} nuclear response function the
upper boundary of the corresponding nuclear physics uncertainty band found in
Sec.~\ref{sec:W}. In Fig.~\ref{fig:He3_shell} relative differences are
moderate:~30\% or less for all recoil directions. The {\it ab initio}
calculation performed here predicts lower rates of recoil events for interaction
operators which generate single nuclear response functions, such as the
$\hat{\mathcal{O}}_{9}$ and $\hat{\mathcal{O}}_{10}$ operators. However, there
are cases in which the {\it ab initio} calculation predicts three nuclear
response functions different from zero, while the phenomenological approach
predicts only two. This is the case of the operator $\hat{\mathcal{O}}_{12}$,
for which {\it ab initio} and phenomenological calculation predict
$W_{k}^{\tau\tau'}\neq 0$ for $k=\Sigma',\Sigma'',\Phi''$ and zero otherwise,
and $W_{k}^{\tau\tau'}\neq 0$ for $k=\Sigma',\Sigma''$ and zero otherwise,
respectively. In these cases, {\it ab initio} rates are larger than
phenomenological rates in a wide range of recoil directions.

\section{Discussion and outlook}
\label{sec:conclusions}

\noindent In this work we have developed an \textit{ab initio} framework for
computations of nuclear response functions for dark matter scattering off atomic
nuclei. Our nuclear-structure calculations have been performed with the NCSM
method, and applied to the study of light nuclei. However, our approach can be
generalized to other \textit{ab initio} methods and extended to heavier
isotopes.

In particular we have quantified the uncertainties of nuclear response functions
that result from the remaining freedom in the construction of realistic nuclear
interactions. Furthermore, we have quantified the impact of such nuclear-physics
uncertainties on physical observables that are relevant for dark matter
searches. Particular emphasis has been placed on the rate of dark
matter--nucleus scattering events at directional detection experiments. We have
performed this calculation for a variegated set of dark matter--nucleon
interactions. Depending on the type of nuclear response that is considered,
relative uncertainties on the scattering rate can be as large as a factor of 5
for nuclear recoil directions antiparallel to the Earth's motion in the galactic
rest frame (see Fig.~\ref{fig:He4_isos}). For comparison, current uncertainties
on the local dark matter density are at the 30\% level~\cite{Pato:2015dua}, or
smaller if knowledge of the baryonic mass density profile is
assumed~\cite{Catena:2009mf}. Uncertainties on the local dark matter velocity
distribution can be significantly larger, but only affect results obtained for
dark matter particle masses below 20 GeV or so~\cite{Catena:2011kv}. We have
also compared scattering rates computed using the NI-SM with those from the
\textit{ab initio} NCSM approach. For \nuc{3,4}{He} most differences are
moderate or small, although with a clear dependence on the nuclear recoil
direction. However, certain response functions that are evaluated to zero in the
NI-SM approach can appear when allowing more freedom in the nuclear many-body
model space. Consequently, we have identified scenarios in which the expected
dark matter signal is larger when \textit{ab initio} nuclear-structure input is
used.

In this work we have used the NREFT description of the WIMP--nucleon
interaction. However, it should be straightforward to implement also the
alternative framework in which QCD constraints, imposed by chiral symmetry, are
used to obtain the WIMP--nucleon interaction from an underlying interaction at
the quark level. Such an extension would become relevant for the matching of
parameters to new physics models, but also for an improved understanding of the
relative importance of many-body currents in nuclei with larger mass numbers.

Further applications of the \textit{ab initio} scheme that we have developed
include improved calculations of (i) the rate of dark matter capture via
scattering by nuclei in the Earth and in the Sun; (ii) the nuclear response
functions for \nuc{19}{F}, which is used in the direct detection experiment
PICO~\cite{Amole:2016pye} and in directional detection
experiments~\cite{Mayet:2016zxu}; and (iii) the nuclear response functions for
\nuc{16}{O}, that is used in the direct detection experiment
CRESST-II~\cite{Angloher:2015ewa}. In order to maintain reasonable
nuclear-physics uncertainties for predictions involving these heavier isotopes
one might have to calibrate the chiral nuclear interaction differently from what
has been done in this work. In particular, rather than staying exclusively in
the few-body sector one might explore the alternative strategy of informing the
nuclear-force model about low-energy many-body
observables~\cite{PhysRevC.91.051301,Hagen:2015yea}. In addition, one could
consider to include information on various electroweak observables, which would
provide additional constraints on the relevant response functions.

Furthermore, medium-mass and heavier closed-shell nuclei can be within
computational reach using nuclear structure methods that have a gentler scaling
with the number of nucleons. In addition, such methods can be used to compute
effective valence-space interactions for use in standard SM
calculations~\cite{PhysRevLett.113.142501,PhysRevLett.113.142502}. A key point
would be that such valence-space interactions will be constructed directly from
the underlying \NN\ interaction using nonperturbative methods. With
corresponding advances in SM technology, this approach opens the path towards
\textit{ab initio} studies of nuclear responses for germanium and xenon isotopes
with quantified uncertainties.

\begin{acknowledgments}
\noindent
This work has been supported by the Knut and Alice Wallenberg Foundation (PI:
Jan Conrad) and is performed in the context of the Swedish Consortium for Direct
Detection of Dark Matter (SweDCube). This research was also supported by the
Munich Institute for Astro- and Particle Physics (MIAPP) of the DFG cluster of
excellence ``Origin and Structure of the Universe'' and we thank the
participants of the program on ``Astro-, Particle and Nuclear Physics of Dark
Matter Direct Detection'' for many valuable discussions. Some of the
computations were performed on resources provided by the Swedish National
Infrastructure for Computing (SNIC) at NSC.
\end{acknowledgments}

\appendix*
\section{Dark matter response functions}
\label{sec:appDM}
\noindent Dark matter response functions introduced in Eq.~(\ref{eq:dsigma}) and
used in Sec.~\ref{sec:dmpheno}: 
\begin{align} \label{eq:R}
R_{M}^{\tau \tau^\prime}\left(v_{q}^{\perp 2}, {q^2 \over m_N^2}\right) &= c_1^\tau c_1^{\tau^\prime } + {J_\chi (J_\chi+1) \over 3} \left( {q^2 \over m_N^2} v_{q}^{\perp 2} c_5^\tau c_5^{\tau^\prime }\right. \nonumber \\ 
 &+ \left. v_{q}^{\perp 2}c_8^\tau c_8^{\tau^\prime} + {q^2 \over m_N^2} c_{11}^\tau c_{11}^{\tau^\prime } \right), \nonumber \\
 R_{\Phi^{\prime \prime}}^{\tau \tau^\prime}\left(v_{q}^{\perp 2}, {q^2 \over m_N^2}\right) &= {q^2 \over 4 m_N^2} c_3^\tau c_3^{\tau^\prime } + {J_\chi (J_\chi+1) \over 12} \nonumber \\
&\times   \left( c_{12}^\tau-{q^2 \over m_N^2} c_{15}^\tau\right)\hspace{-0.1 cm}\left( c_{12}^{\tau^\prime }-{q^2 \over m_N^2}c_{15}^{\tau^\prime} \right),  \nonumber \\
 R_{\Phi^{\prime \prime} M}^{\tau \tau^\prime}\left(v_{q}^{\perp 2}, {q^2 \over m_N^2}\right) &=  c_3^\tau c_1^{\tau^\prime } + {J_\chi (J_\chi+1) \over 3} \nonumber \\
&\times \left( c_{12}^\tau -{q^2 \over m_N^2} c_{15}^\tau \right) c_{11}^{\tau^\prime }, \nonumber \\
R_{\tilde{\Phi}^\prime}^{\tau \tau^\prime}\left(v_{q}^{\perp 2}, {q^2 \over m_N^2}\right) &={J_\chi (J_\chi+1) \over 12} \nonumber \\
&\times \left( c_{12}^\tau c_{12}^{\tau^\prime }+{q^2 \over m_N^2}  c_{13}^\tau c_{13}^{\tau^\prime}  \right), \nonumber \\
R_{\Sigma^{\prime \prime}}^{\tau \tau^\prime}\left(v_{q}^{\perp 2}, {q^2 \over m_N^2}\right)  &={q^2 \over 4 m_N^2} c_{10}^\tau  c_{10}^{\tau^\prime } + {J_\chi (J_\chi+1) \over 12} \Bigg[ c_4^\tau c_4^{\tau^\prime}   \nonumber \\
&+ {q^2 \over m_N^2} ( c_4^\tau c_6^{\tau^\prime }+c_6^\tau c_4^{\tau^\prime }) + {q^4 \over m_N^4} c_{6}^\tau c_{6}^{\tau^\prime } \nonumber \\
&+v_{q}^{\perp 2} c_{12}^\tau c_{12}^{\tau^\prime }+{q^2 \over m_N^2} v_{q}^{\perp 2} c_{13}^\tau c_{13}^{\tau^\prime } \Bigg], \nonumber \\
R_{\Sigma^\prime}^{\tau \tau^\prime}\left(v_{q}^{\perp 2}, {q^2 \over m_N^2}\right)  &={1 \over 8} \left( {q^2 \over  m_N^2}  v_{q}^{\perp 2} c_{3}^\tau  c_{3}^{\tau^\prime } + v_{q}^{\perp 2}  c_{7}^\tau  c_{7}^{\tau^\prime }  \right) \nonumber \\
&+ {J_\chi (J_\chi+1) \over 12} \Bigg[ c_4^\tau c_4^{\tau^\prime} + {q^2 \over m_N^2} c_9^\tau c_9^{\tau^\prime }  \nonumber \\
&+\hspace{-0.1 cm}{v_{q}^{\perp 2} \over 2} \hspace{-0.1 cm} \left(\hspace{-0.05 cm}c_{12}^\tau-{q^2 \over m_N^2}c_{15}^\tau \hspace{-0.05 cm}\right) \hspace{-0.1 cm}\left( \hspace{-0.05 cm}c_{12}^{\tau^\prime }-{q^2 \over m_N^2}c_{15}^{\tau \prime} \hspace{-0.05 cm}\right) \nonumber \\
&+ {q^2 \over 2 m_N^2} v_{q}^{\perp 2} c_{14}^\tau c_{14}^{\tau^\prime } \Bigg], \nonumber \\
R_{\Delta}^{\tau \tau^\prime}\left(v_{q}^{\perp 2}, {q^2 \over m_N^2}\right)&=  {J_\chi (J_\chi+1) \over 3} \left( {q^2 \over m_N^2} c_{5}^\tau c_{5}^{\tau^\prime }+ c_{8}^\tau c_{8}^{\tau^\prime } \right), \nonumber \\
R_{\Delta \Sigma^\prime}^{\tau \tau^\prime}\left(v_{q}^{\perp 2}, {q^2 \over m_N^2}\right)&= {J_\chi (J_\chi+1) \over 3} \left(c_{5}^\tau c_{4}^{\tau^\prime }-c_8^\tau c_9^{\tau^\prime} \right).
\end{align}
For definitiveness, we assume $J_\chi=1/2$, where $J_\chi$ is the dark matter
particle spin.

\nocite{apsrev41Control}
\bibliography{revtex_control,refs-DM-abinitio,refs-DM-abinitio_2}

\end{document}